\documentclass[12pt]{article}
\usepackage{graphicx}
\usepackage{dcolumn}
\usepackage{bm}
\usepackage{color}
\usepackage{amssymb,amsmath,amsfonts}

\def\Title#1{\begin{center} {\Large #1 } \end{center}}
\def\Author#1{\begin{center}{ \sc #1} \end{center}}
\def\Address#1{\begin{center}{ \it #1} \end{center}}

\newenvironment{Abstract}{\begin{quotation}  }{\end{quotation}}
\def\Acknowledgements{\bigskip  \bigskip  \begin{large}
             \bf Acknowledgements \end{large}}

\begin{document}

\Title{Residence and Waiting Times of Brownian Interface Fluctuations}

\Author{V.W.A. de Villeneuve$^{1}$, J.M.L. van Leeuwen$^{2}$,  J.W.J. de Folter$^{1}$, D.G.A.L. Aarts$^{3}$, W. van Saarloos$^{2}$ and H.N.W. Lekkerkerker$^{1}$}
\Address{$^{1}$ Van 't Hoff Laboratory for Physical and Colloid Chemistry, University of Utrecht - Padualaan 8, 3584 CH Utrecht, The Netherlands\\ $^{2}$Instituut-Lorentz for Theoretical Physics, Leiden University, Niels Bohrweg 2, Leiden, 2333 CA, The Netherlands\\
$^{3}$Physical and Theoretical Chemistry Laboratory, University of Oxford, South Parks Road, Oxford OX1 3QZ, UK\\}
\begin{Abstract}
We report on the residence times of capillary waves above a given height $h$ and on the typical waiting time in between such fluctuations. The measurements were made on phase separated colloid-polymer systems by laser scanning confocal microscopy. Due to the Brownian character of the process, the stochastics vary with the chosen measurement interval $\Delta t$. In experiments, the discrete scanning times are a practical cutoff and we are able to measure the waiting time as a function of this cutoff. The measurement interval dependence of the observed waiting and residence times turns out to be solely determined by the time dependent height-height correlation function $g(t)$. We find excellent agreement with the theory presented here along with the experiments.\end{Abstract}

\section{Introduction}
The often counterintuitive field of stochastic number fluctuations has boggled the scientist's mind for ages.
Svedberg's measurement of the diffusion constant \cite{svedberg11}, based on the residence times on a fluctuating number of Brownian particles in a fixed region, inspired von Smoluchowski to develop a theoretical framework for it \cite{smoluchowski14,smoluchowski15}.
His ideas were for example exploited to derive the mobility of spermatozoids \cite{rothschild1953} and white blood cells \cite{nossal1976}. The original experiment was performed in even more detail by Brenner, Weiss and Nossal \cite{brenner1978}.
Based on Einstein's \cite{einstein05} and Perrin's seminal papers on Brownian motion of particles \cite{perrin08}, von Smoluchowski \cite{Smoluchowski08} was the first to predict the Brownian height fluctuations of the interface.
They were first theoretically treated by Mandelstam \cite{mandelstam13} and have become an important component of modern theories of interfaces \cite{buff65, mecke99,milchev02}.
Capillary waves were initially accessed experimentally by light-\cite{vrij68} and x-ray scattering \cite{fradin00,sanyal91}.
On a microscopic level capillary waves were studied in computer simulations of molecular systems  \cite{sikkenk87}, before recent investigations by Aarts and coworkers on colloid-polymer mixtures, \cite{Aarts2004science,derks2006,royall2007} added another dimension to studies on capillary waves by using confocal microscopy. 
In these experiments, the interfacial tension $\gamma$ is lowered to the $nN/m$ range.
As a consequence the characteristic length and time scale of the fluctuations are such that they can be visualized by microscopy.
Microscopy furthermore enables the investigation of local phenomena such as the role of capillary waves in the rupture \cite{hennequin2006} and coalescence \cite{aarts2007sm} problems.
At a fixed location on the interface, the height is continuously fluctuating in time. 
Following Becker \cite{becker66} and as sketched out in Fig.\ref{figcartoon}A the waiting time $\Theta$ is the average time spent in between fluctuations above a height $h$, and the residence time $T$ is the average time spent above a height $h$:
\begin{figure}
\includegraphics[width=12 cm]{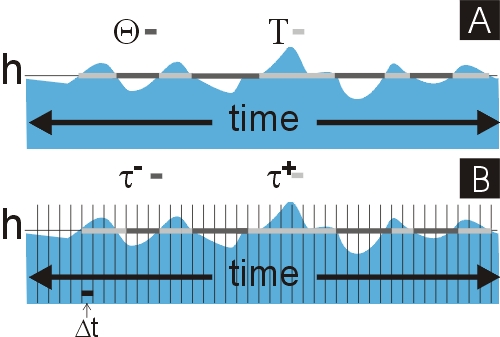}
\centering
\caption{(A)Variation of the interfacial height $h$ in time with characteristic residence times $T$ and waiting times $\Theta$. (B) in practice, $T$ and $\Theta$ are determined by discrete intervals, resulting in observed residence and waiting times $\tau^+$ and $\tau^-$. These are determined by weighted averages, of the stochastics of intervals with length $\Delta t$.}
\label{figcartoon}
\end{figure}
\begin{equation}\label{eqwaiting}
\Theta(h)= \int dt p^- (h,t) t, \quad  \quad  T(h)= \int dt p^+(h,t) t
\end{equation}
where $p^-(h,t)$ and $p^+(h,t)$ are the probabilities on intervals of length $t$ below and above the height $h$ respectively.
Such local measurements are not possible by scattering methods, since they require knowledge of continuous local stochastics which seem, however,  easily accessible by microscopy.
Pioneering work by Aarts and Lekkerkerker \cite{AartsJFM} resulted in scaling relations, but did not include a full quantitative description of the process.
The crux is that however fast microscopy may be, measurements always have to be taken at discrete time intervals $\Delta t$, as sketched in Fig. \ref{figcartoon}B, resulting in the observed waiting and residence times $\tau^+$ and $\tau^-$:
\begin{equation}\label{eqwaitingapp}
\tau^-(h)=\sum n p^-_n (h), \quad  \quad  \tau^+(h)=\sum n p^+_n (h) 
\end{equation}
Here, $p^-_n$ and $p^+_n$ are respectively the (normalized) probabilities on intervals of snapshots of length $n$, below and above height $h$. 
Note that $\tau^+$ and $\tau^-$ are in units $\Delta t$ and are therefore both dimensionless.
Switching to discrete time intervals is not entirely trivial. Due to the Brownian character of the process, the discretisation of \ref{eqwaiting} leads to statistics that depend on the chosen interval $\Delta t$.
However, the necessity to be discrete saves rather than spoils the day, as it enables us to overcome the divergencies for continuous distributions.

\begin{figure}
\includegraphics[width=12 cm]{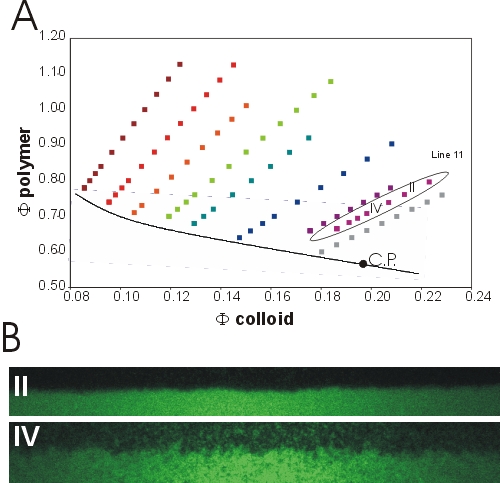}
\centering
\caption{(A) Phase diagram of the system studied. The dilution line (line 11) studied here is marked. All shown state points phase separate. The indicated binodal is a guide to the eye: the theoretical binodal occurs at much lower concentrations. The marked critical point is an estimate based on the ratios of the volumes of the "liquid" and "gas" phases. (B) Confocal Images of Statepoint II and IV.}
\label{phase diagram}
\end{figure}
 
\section{Experimental Section}
Fluorescently labeled polymethyl-metacrylate particles were prepared using the Bosma method \cite{Bosma2002}, slightly modified by using decalin (Merck, for synthesis) as the reaction solvent \cite{aartscodef}.
The particle polydispersity is around 10\% from scanning electron microscopy, and the dynamic light scattering particle radius is 62 nm.
Polystyrene (2 10$^3$ kg mol$^{-1}$) was added as depletant polymer, with an estimated radius of gyration $R_g$ of 42 nm \cite{vincent90}.
At sufficiently high colloid and polymer volume fractions, respectively $\phi_c = \pi/6 \sigma_c^3 n_c$ and $\phi_p = 4/3 \pi R_g^3 n_p$ (with $n_c$ and $n_p$ the number densities of colloids and polymers respectively), this system phase separates into a colloid-rich (colloidal liquid) and a polymer rich (colloidal gas) phase.
By diluting several phase separating samples with its solvent decalin, the phase diagram presented in Fig. \ref{phase diagram}A was constructed. The shown binodal is a guide to the eye (the theoretical binodal appears at much lower volume fractions) and the critical point is an estimate based on the ratios of the volumes of the phases.

A Nikon Eclipse E400 laser scanning confocal microscope equipped with a Nikon C1 scanhead was placed horizontally to study the colloid polymer mixture \cite{aartscodef}.
The microscope was furthermore equipped with a 405 nm laser and a Nikon 60X CFI Plan Apochromat (NA 1.4) Lens.
The sample container is a small glass vial, part of which is removed and replaced by a thin (0.17 mm) glass wall.
Series of 10 000 snapshots of the interface of 640 x 64 and 640 x 80 pixels (100 $\times$ 10.0 and 12.5 $\mu m$) were taken at constant intervals $t_i$ of 0.45 $s$ and 0.50 $s$ of statepoints II and IV along the marked dilution line in the phase diagram, shown in Fig. \ref{phase diagram}.
A single scan takes approximately 0.25 s to complete (the exact scan time does vary by a few percent in time).
Typical snapshots are shown for a number of state points in Fig. \ref{phase diagram}b.
The low excitation wavelength results in resolution of $\sim$ 160 nm. The particles are $\sim$ 138 nm in diameter, hence a pixel roughly corresponds to a particle.
Note that the resolution of the measured heights is significantly higher than this:
the vertical location of the interface $h(x)$ is determined for each column of pixels in a frame by fitting the pixel value $I(z)$, which is proportional to the local colloid concentration, to a van der Waals profile $I(z) = a + b$ tanh$([z-h(x)]/c)$, as in \cite{derks2006}.
The exact resolution depends on the contrast between the phases, which depends on the distance to the critical point.
Further away from the binodal the height fluctuations are too small, closer to the binodal the contrast between the phases vanishes and the capillary waves start to show overhang effects.

\begin{figure}
\includegraphics[width=12 cm]{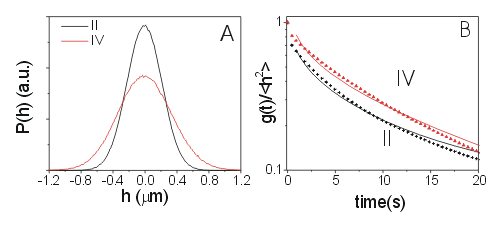}
\centering
\caption{Interfacial properties determined by dynamic correlation functions and height distributions for state points II and IV. (A) height distributions for several state points. (B) The dynamic correlation functions obtained for statepoints II and IV, normalized by $\langle h^2\rangle$.}
\label{variables}
\end{figure}

\section{Observations and Results}
The wave heights $h$ at an arbitrary point are distributed according to the Gaussian
\begin{equation} \label{g2}
P_{eq} (h) = {\exp (-h^2 /2 \langle h^2 \rangle) \over [2 \pi \langle h^2 \rangle]^{1/ 2}}.
\end{equation} 
It can be shown that $\langle h^2 \rangle$ satisfies the expression
\begin{equation} \label{heights}
\langle h^2 \rangle = \frac{L^2}{2\pi}\int_{k_{min}}^{k_{max}} \langle |h_k |^2 \rangle k dk = \frac{k_B T}{4\pi \gamma}\ln \frac{k^2_{max}+\xi^{-2}}{k^2_{min}+\xi^{-2}}.
\end{equation} 
Here $\xi=(\gamma /\Delta \rho g)^{1/2}$ is the the capillary length, with $g$ the gravitational acceleration, $\Delta\rho$ the density difference between the two phases, $k_B$ the Boltzmann constant, $T$ the Kelvin temperature and $L$ the system size.
Furthermore, $h_k$ is the amplitude of mode $k$, $k_{max}\approx 2\pi/d$ is a cutoff related to the typical interparticle distance $d$ and $k_{min}=2\pi/L$ is related to the system size $L$.
In equation (\ref{heights}) we have used that the mean square average of $h_k$ is given by
\begin{equation} \label{average}
\langle |h_k |^2 \rangle = {k_B T \over L^2(\Delta\rho g + \gamma k^2)}.
\end{equation} 
The height distributions are shown for statepoints II and IV in Fig. \ref{variables}A, with  $\langle 
h^2\rangle^{1/2} = 0.219$ and $0.336 \mu m$ for statepoints II an IV respectively.
These will be used as a unit for the heights.

Next we consider the height correlations in time $\langle h(t)h(t')\rangle$ at a fixed position, which we denote by $\langle h^2 \rangle g(t-t')$.
It is calculated as: \cite{jeng98,meunier88}
\begin{equation} \label{eqtcorr}
\langle h^2 \rangle g(t-t') = \frac{L^2}{2\pi}\int_{k_{min}}^{k_{max}} \langle |h_{k} |^2 \rangle  e^{-\omega_k |t-t'|}kdk.
\end{equation}
The measured and fitted (setting $k_{max}=\infty$ and $k_{min}=0$) dynamic correlation functions of statepoints II and IV are shown in Fig. \ref{variables}B.
In colloidal systems capillary waves are in the overdamped regime \cite{jeng98,Aarts2004science} with a decay rate
\begin{equation} \label{dampfreq}
\omega_k = {1 \over 2 t_c}  (k\xi+(k\xi)^{-1}).
\end{equation} 
Here, $t_c= \xi /u_c$ is the capillary time and $u_c=\gamma/\eta$ is the capillary velocity with $\eta$ the combined viscosities of the phases.
From the fits, the interfacial tensions and the capillary times can be extracted, which gives $t_c = 12s$ and $\gamma = 69 nN/m$ for statepoint II and $t_c = 22s$ and $\gamma = 25 nN/m$ for statepoint IV.

\begin{figure}
\includegraphics[width=12 cm]{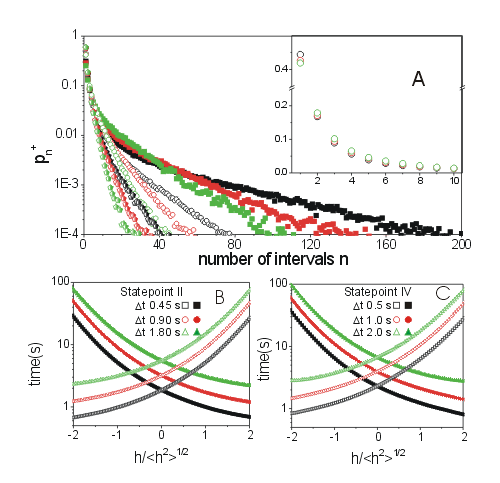}
\centering
\caption{(A) The distribution of observed residence times $\tau^+$ for statepoint IV. $p^+_n(h)$ is shown for h = -1 (filled squares), 0 (circles), and 1 (semi-filled pentagons) $\times$ $\langle h^2 \rangle^{{1 \over 2}}$.  The inset shows the short time behaviour for $h=0$ in detail. The colours correspond to time intervals $\Delta$ t = 1 $t_i$ (black squares), 2 $t_i$ (red circles) and 4 $t_i$ (green triangles). (B) Observed waiting times (open symbols) and residence times (filled symbols), observed for statepoint II, at time intervals of 1 $t_i$ (black squares), 2 $t_i$ (red circles) and 4 $t_i$ (green triangles). (C) as in panel (B), but now for statepoint IV.}
\label{TimesI}
\end{figure}

The residence and waiting times are calculated through the experimentally obtained functions $p^\pm_n (h)$. In order to check the possible variation of waiting and residence times, we calculate the distributions for intervals of 1,2 and 4 $t_i$. They are shown for statepoint IV at heights $h=-1,0$ and $1 \langle h^2\rangle^{1/ 2}$ in Fig. \ref{TimesI}A. Note that the x-axis has intervals as units, not time. The distributions are clearly complicated:
the shortest interval of 1 time interval becomes more dominant as $\Delta t$ decreases as shown in the inset of Fig. \ref{TimesI}A.
On the other hand, $h=-\langle h^2\rangle^{1\over 2}$ has a surprisingly large interval of $\approx$ 2-50 interval durations, which occur most frequently for $\Delta t = 4 t_i$.
The longest intervals decay exponentially and as expected decrease most rapidly in quantity for the largest time intervals.
As a result, the waiting and residence times, shown for statepoints II and IV  as a function of height in Fig. \ref{TimesI}B and Fig. \ref{TimesI}C, are confusing at first sight:
if we increase the time in between observations $\Delta t$ to 2 or 4 $t_i$, the observed residence times are significantly larger!
\section{Interpretation and Discussion}
In order to understand these observations, we need some theoretical framework. 
In the spirit of von Smoluchowski we start with simple counting arguments, 
\cite{smoluchowski14,smoluchowski15}, in order to identify the essential 
object to be calculated. Consider a long measurement of $N$ snapshots. 
We take pairs of consecutive heights and
divide them into 4 sets: $N^{++} (h)$ pairs where on both sides
of the interval the interface is above $h$, $N^{+-} (h)$ pairs where the 
earlier value is above $h$ and the later value below $h$, 
$N^{-+} (h)$ pairs where it is the other way around and finally 
$N^{--} (h)$  pairs where the interface is on both sides of the interval 
below $h$. The first observation is that 
\begin{equation} \label{h0}
N^{+-} (h) =N^{-+} (h) \equiv M(h),
\end{equation} 
since each interval in which the interface crosses 
the level $h$ from above is followed by the next crossing from below. 
These numbers are also equal to the number of $M(h)$ of ``hills'', which in 
turn is the same as the number of ``valleys''. 
Now $M (h)\tau^+(h)$ is the total length of the hills (measured in units 
$\Delta t$). Thus we have the relations
\begin{equation} \label{h1}
M(h) \tau^+(h) = N^{++} (h)+ M(h), M(h) \tau^-(h) = N^{--}(h) + M (h).
\end{equation} 
We have to add $M (h)$ on the right hand side, since the number of points in a 
hill (valley) is one more than the number of intervals inside a hill (valley).  
Adding the two relations gives
\begin{equation} \label{h2} 
\tau^+ (h)+ \tau^- (h)= {N \over M(h)} \equiv {1 \over r(h)}.
\end{equation}
$r(h)$ is the probability to find an interval where the interface crosses the
level $h$ from above to below $h$.

\begin{figure}
\includegraphics[width=12 cm]{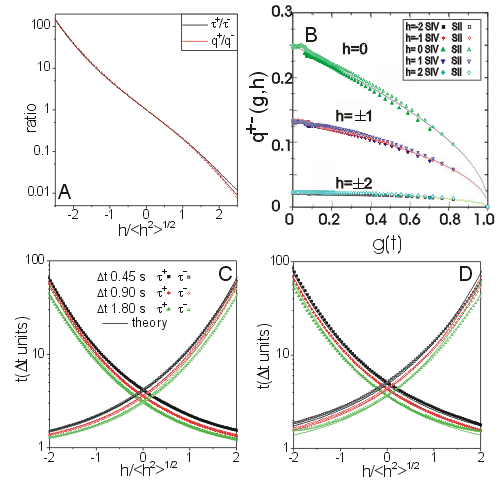}
\centering
\caption{Waiting and residence times for statepoints II and IV. (A) The relation $q^+(h)/q^-(h)=\tau^+(h)/\tau^-(h)$ clearly holds over many orders of magnitude, both for statepoint II (dashed lines) and IV (non-dashed lines). (B) the probability $r(h,g)$ as a function of $g$ for $|h|$ = 0, 1 and 2 $\langle h^2 \rangle^{1/ 2}$. The lines are the theoretical expressions, the points are obtained through the definition of $r$ from the experiment. (C) Theoretical and observed time intervals for the waiting and residence times of statepoint II, with $\Delta t$ = 1 (black squares), 2 (red circles) and 4 (green triangles) $t_i$. Open symbols are waiting times, closed symbols are residence times. The lines are the theoretical curves. (D) Theoretical and observed time intervals for statepoint IV, with $\Delta t$ = 1,2 and 4 $t_i$. Symbols are identical to those in panel (C).}
\label{waitingtimes}
\end{figure}

The second observation is that the right hand side of (\ref{h1}) gives the 
probability to find a point above (below) $h$:
\begin{equation} \label{h3}
{N^{++} (h) + M(h) \over N} = q^+ (h) =  \int^\infty_h dh' P_{eq} (h'), {N^{--} (h) + M(h) \over N} = q^- (h) =  \int^h_{-\infty} dh' P_{eq} (h'),
\end{equation} 
which can be calculated from the equilibrium distribution and which states that the fraction of heights above or below $h$ is equivalent to the fraction of time spent above or below height $h$.
Thus we find the expressions for $\tau^{\pm} (h)$
\begin{equation} \label{h4}
\tau^{\pm} (h) = {q^{\pm} (h) \over r(h)},
\end{equation} 
showing that $r(h)$ is the basic quantity to be calculated. 
Equation (\ref{h4}), independently of $r(h)$, implies:
\begin{equation}\label{eqfrac}
\frac{\tau^+(h)}{\tau^-(h)}=\frac{q^+(h)}{q^-(h)},
\end{equation}
which is experimentally verifiable. In Fig. \ref{waitingtimes}A we show that both sides of \ref{eqfrac} fall on a single mastercurve over nearly the full spectrum of heights for both statepoints.

For $r(h)$ we need the joint probability to find the interface on level $h'$ at
$t=0$ and on $h''$ at time $t$ 
\begin{equation} \label{d4}
\langle \delta(h(0,0) - h') \delta (h(0,t) - h'') \rangle = P_{eq} (h')  G_c (h',0| h'', t).
\end{equation}
It is the product of the height distribution function and the conditional 
probability, that starting from $h'$ at $t=0$ one arrives at $h''$ a
time $t$ later. $G_c$ follows from the Langevin equation for the interface 
modes $k$, using a fluctuation force with a white noise spectrum \cite{Kampen}.
The result is
\begin{equation} \label{e3}
G_c( h',0 | h'',t ) = {1 \over [2 \pi \langle h^2 \rangle (1 -g^2(t))]^{\frac{1}{2}}}\exp{- {[h'' - h' g(t)]^2 \over 2 \langle h^2 \rangle [1-g^2 (t)]}}.
\end{equation}
We point out that $G_c$ contains only the correlation function $g(t)$ and not 
explicitely the decay rate $\omega_k$. A check on (\ref{e3}) is that the 
correlation function $\langle h(0) h(t) \rangle$ as calculated from 
(\ref{d4}) and (\ref{e3}) gives the result $\langle h^2 \rangle g(t)$.
Fig. \ref{variables}B shows that $g(t)$ is not exponentially decaying.
Thus the spatial process is non-Markovian \cite{Kampen}.

According to \ref{d4} $r(h)$ is given by the integral
\begin{equation} \label{g1}
r(h, \Delta t) = \int^\infty_h dh' P_{eq} (h') \int^h_{-\infty}  dh''G_c (h',0| h'',\Delta t),
\end{equation} 
where the first integral selects the points above $h$ with the height 
distribution and the second integral selects the points below $h$ with the 
conditional probability given by (\ref{e3}).
We have made the dependence on the interval length $\Delta t$ explicit 
because it is an important issue. $\Delta t$ enters in the calculation of 
$r(h, \Delta t)$ through the value $g(\Delta t)$. We may therefore consider 
$r$ also as a function of $h$ and $g=g(\Delta t)$, using the experimental 
values of $g(t)$ as input. Of course $r(h,g)$ is experimentally measurable 
using its definition (\ref{h2}).
The theoretical and experimental $r(h,g)$ are plotted in Fig. \ref{waitingtimes}B as function 
of $g$ for a few values of $h$ for both statepoints, resulting in a single curve for each $h/\langle h^2\rangle\frac{1}{2}$. The agreement between theory and experiment is remarkable. The observed residence and waiting times for $\Delta t = 1, 2$ and $4 t_i$ for 
statepoints II and IV, are shown in Fig. \ref{waitingtimes}C and \ref{waitingtimes}D in terms of 
units $\Delta t$. Again we find excellent agreement with the theory, using the measured values for the 
$g(t)$ function as theoretical input for the determination of $r$. Note that 
the y-axis has units "$\Delta t$" and is dimensionless. In terms of time, the 
residence time at height $h$ increases with $\Delta t$.  At sufficiently large 
$h$, the theoretical times are always larger than the experimental values,
see Fig. \ref{waitingtimes}C and \ref{waitingtimes}D.
This is at least partly due to the limited amount of time points considered, 
which tends to exclude the tail of the distributions $p^\pm_n(h)$.

We now ask ourselves the question whether we can take the limit of 
$\Delta t \rightarrow 0$ and arrive at a continuum description. If 
$r(h, \Delta t)$ were to shrink proportional to $\Delta t$, the $\tau^\pm (h)$ 
would increase inversely proportional to $\Delta t$. From a continuum
description one would expect the relations
\begin{equation} \label{g7}
T = \lim_{\Delta t \rightarrow 0} \tau^+ (h) \Delta t, \quad \quad \Theta = \lim_{\Delta t \rightarrow 0} \tau^- (h) \Delta t
\end{equation} 
and $T$ and $\Theta$ would have a finite limit.
However, analyzing the small $\Delta t$ behavior of the expression 
(\ref{g1}), which amounts to the limiting behavior for 
$g(\Delta t) \rightarrow 1$, we find
\begin{equation} \label{g8}
r(h,g \rightarrow 1) \simeq {\sqrt{1-g} \over \pi \sqrt{2} } \exp \left
[- {h^2  \over 2 \langle h^2 \rangle}\right].
\end{equation}
The square root in (\ref{g8}) is a reflection of the Brownian character of the fluctuations.
If the initial decay of $g(t)$ is linear, $r(h,g)$ vanishes as $(\Delta t)^{1/2}$ and consequently $T$ and $\Theta$ vanish.

With \ref{h4} and the computation of $r(h,g)$ we have determined the
residence and waiting times without evaluating the distributions
$p^\pm_n (h)$. The latter quantities can also be calculated using multiple
correlation functions, which are generalizations of \ref{d4}. Their
evaluation is increasingly involved, due to the non-Markovian character
of the spatial process. Experimentally, the exhaustive amount of
multiple correlation functions can be determined given sufficient time points, analogous to the results in Fig. \ref{waitingtimes}B. We leave this aspect for a further study.

\section{Conclusion}
We have presented confocal microscopy experiments along with theory for the microscopic waiting and residence times of heights $h$ of the capillary waves of the fluid-fluid interface of a phase separated colloid-polymer mixture.
Due to the Brownian character of the process, these times depend on the experimental measurement interval $\Delta t$. The results from this discrete time sampling are predictable in terms of  the decay of the height-height correlation function $g(t)$, which in principle includes the relevant system variables interfacial tension and capillary time. We find excellent agreement between experiments and theory.

\Acknowledgements
The authors are indebted for valuable discussions to H.W.J. Bl\"ote and W.Th.F. den  Hollander. We additionally thank C. Vonk for assistance with particle synthesis, S. Sacanna for help with the electron microscopy and B. Kuipers for aid with the dynamic light scattering experiments.
The work of VWAdV is part of the research program of the 'Stichting voor Fundamenteel Onderzoek der Materie (FOM)', which is financially supported by the 'Nederlandse Organisatie voor Wetenschappelijk Onderzoek (NWO)'.
Support of VWAdV by the DFG through the SFB TR6 is acknowledged.
\bibliographystyle{unsrt}
\bibliography{libold} 
\end{document}